\documentclass[reqno,openright,dvips,11pt,myheadings,twoside]{amsart}

\newtheorem{theorem}{Theorem}[section]
\newtheorem{lemma}[theorem]{Lemma}
\newtheorem{cor}[theorem]{Corollary}

\theoremstyle{definition}
\newtheorem{definition}[theorem]{Definition}

\theoremstyle{remark}

\numberwithin{equation}{section}

\newcommand{\abs}[1]{\lvert#1\rvert}

\newcommand{\BB}[1]{\ensuremath{\mathbb{#1}}}
\newcommand{\R}{\ensuremath{\BB{R}}}
\newcommand{\iny}{\ensuremath{\infty}}
\newcommand{\grad}{\ensuremath{\nabla}}
\DeclareMathOperator{\dv}{div}
\newcommand{\prt}{\ensuremath{\partial}}
\newcommand{\brac}[1]{\ensuremath{\left[ #1 \right]}}
\newcommand{\pr}[1]{\ensuremath{\left( #1 \right) }}
\newcommand{\set}[1]{\ensuremath{\left\{ #1 \right\}}}
\newcommand{\norm}[1]{\ensuremath{\left\Vert #1 \right\Vert}}
\newcommand{\smallnorm}[1]{\ensuremath{\Vert #1 \Vert}}
\newcommand{\refS}[1]{Section~\ref{S:#1}}
\newcommand{\refT}[1]{Theorem~\ref{T:#1}}
\newcommand{\refL}[1]{Lemma~\ref{L:#1}}
\newcommand{\refD}[1]{Definition~\ref{D:#1}}
\newcommand{\refC}[1]{Corollary~\ref{C:#1}}
\newcommand{\refE}[1]{Equation~(\ref{e:#1})}
\newcommand{\eps}{\ensuremath{\epsilon}}
\newcommand{\Cal}[1]{\ensuremath{\mathcal{#1}}}
\newcommand{\al}{\ensuremath{\alpha}}
\newcommand{\la}{\ensuremath{\lambda}}
\newcommand{\pdx}[2]{\frac{\prt #1}{\prt #2}}
\newcommand{\diff}[2]{\frac{ d#1}{d#2}}
\newcommand{\ol}{\overline}
\newcommand{\smallabs}[1]{\ensuremath{\vert #1 \vert}}

\begin{document}

\raggedbottom

\numberwithin{equation}{section}

%
%
\newcommand{\MarginNote}[1]{
    \marginpar{
        \begin{flushleft}
            \footnotesize #1
        \end{flushleft}
        }
    }

%
%
\newcommand{\NoteToSelf}[1]{
    }

\newcommand{\Detail}[1]{
    \MarginNote{Detail}
    \skipline
    \hspace{+0.25in}\fbox{\parbox{4.25in}{\small #1}}
    \skipline
    }

\newcommand{\Todo}[1]{
    \skipline \noindent \textbf{TODO}:
    #1
    \skipline
    }

\newcommand{\Comment}[1] {
    \skipline
    \hspace{+0.25in}\fbox{\parbox{4.25in}{\small \textbf{Comment}: #1}}
    \skipline
    }

%
%

\newcommand{\IntTR}
    {\int_{t_0}^{t_1} \int_{\R^d}}

\newcommand{\IntAll}
    {\int_{-\iny}^\iny}

\newcommand{\Schwartz}
    {\ensuremath \Cal{S}}

\newcommand{\SchwartzR}
    {\ensuremath \Schwartz (\R)}

\newcommand{\SchwartzRd}
    {\ensuremath \Schwartz (\R^d)}

\newcommand{\SchwartzDual}
    {\ensuremath \Cal{S}'}

\newcommand{\SchwartzRDual}
    {\ensuremath \Schwartz' (\R)}

\newcommand{\SchwartzRdDual}
    {\ensuremath \Schwartz' (\R^d)}

\newcommand{\HSNorm}[1]
    {\norm{#1}_{H^s(\R^2)}}

\newcommand{\HSNormA}[2]
    {\norm{#1}_{H^{#2}(\R^2)}}

\newcommand{\Holder}
    {H\"{o}lder }

\newcommand{\Holders}
    {H\"{o}lder's }

\newcommand{\Holderian}
    {H\"{o}lderian }

\newcommand{\HolderRNorm}[1]
    {\widetilde{\Vert}{#1}\Vert_r}

\newcommand{\LInfNorm}[1]
    {\norm{#1}_{L^\iny}}

\newcommand{\SmallLInfNorm}[1]
    {\smallnorm{#1}_{L^\iny}}

\newcommand{\LOneNorm}[1]
    {\norm{#1}_{L^1}}

\newcommand{\SmallLOneNorm}[1]
    {\smallnorm{#1}_{L^1}}

\newcommand{\LTwoNorm}[1]
    {\norm{#1}_{L^2}}

\newcommand{\SmallLTwoNorm}[1]
    {\smallnorm{#1}_{L^2}}

\newcommand{\LpNorm}[2]
    {\norm{#1}_{L^{#2}}}

\newcommand{\SmallLpNorm}[2]
    {\smallnorm{#1}_{L^{#2}}}

\newcommand{\lOneNorm}[1]
    {\norm{#1}_{l^1}}

\newcommand{\lTwoNorm}[1]
    {\norm{#1}_{l^2}}

\newcommand{\MsrNorm}[1]
    {\norm{#1}_{\Cal{M}}}

\newcommand{\FTF}
    {\Cal{F}}

\newcommand{\FTR}
    {\Cal{F}^{-1}}

\newcommand{\InvLaplacian}
    {\ensuremath{\widetilde{\Delta}^{-1}}}

\newcommand{\EqDef}
    {\hspace{0.2em}={\hspace{-1.2em}\raisebox{1.2ex}{\scriptsize def}}\hspace{0.2em}}

%
%
\newcommand{\BoldTau}
    {\mbox{\boldmath $\tau$}}

%
%
\renewcommand{\refE}[1]{(\ref{e:#1})}

\title
    [Navier-Stokes with Navier boundary conditions]
    {Navier-Stokes equations with Navier boundary conditions for a
        bounded domain in the plane}

\author{James P. Kelliher}
\address{Department of Mathematics, University of Texas, Austin,
         Texas, 78712}
\curraddr{Department of Mathematics, University of Texas, Austin,
          Texas, 78712}
\email{kelliher@math.utexas.edu}

\date{} 



\begin{abstract}
    We consider solutions to the Navier-Stokes equations with Navier
    boundary conditions in a bounded domain $\Omega$
    in $\R^2$ with a $C^2$-boundary $\Gamma$. Navier boundary
    conditions can be expressed in the form $\omega(v) = (2 \kappa - \al) v
    \cdot \BoldTau$ and $v \cdot \mathbf{n} = 0$ on $\Gamma$, where $v$
    is the velocity, $\omega(v)$ the vorticity, $\mathbf{n}$ a unit normal
    vector, $\BoldTau$ a unit tangent vector, and $\al$ is in
    $L^\iny(\Gamma)$. Such solutions have been
    considered in \cite{CMR} and \cite{FLP}, and, in the special case where
    $\al = 2\kappa$, by J.L. Lions in \cite{JL1969} and by P.L. Lions in
    \cite{L1996}. We extend the results of \cite{CMR} and \cite{FLP}
    to non-simply connected domains.
    Assuming, as Yudovich
    does in \cite{Y1995}, a particular bound on the growth of the
    $L^p$-norms of the initial vorticity with $p$, and also assuming that
    for some $\eps > 0$, $\Gamma$ is $C^{2, 1/2 + \eps}$ and
    $\al$ is in $H^{1/2 + \eps}(\Gamma) + C^{1/2 + \eps}(\Gamma)$,
    we obtain a bound on the
    rate of convergence in $L^\iny([0, T]; L^2(\Omega) \cap L^2(\Gamma))$
    to the solution to the Euler equations in the vanishing
    viscosity limit. We also show that if
    the initial velocity is in $H^3(\Omega)$ and $\Gamma$ is
    $C^3$, then solutions to the Navier-Stokes equations with Navier
    boundary conditions converge in $L^\iny([0, T]; L^2(\Omega))$
    to the solution to the Navier-Stokes equations with
    the usual no-slip boundary conditions as we let $\al$ grow large
    uniformly on the boundary.
\end{abstract}

\maketitle


%
%
\section{Introduction}\label{S:Introduction}

\noindent Let $\Omega$ be a bounded domain of $\R^2$ with a boundary
$\Gamma$ consisting of a finite number of connected components. We
always assume that $\Gamma$ is at least as smooth as $C^2$, but will
assume additional smoothness as needed.

We consider the existence and uniqueness of a solution $u$ to the
Navier-Stokes equations under \textit{Navier boundary conditions};
namely,
\begin{align}\label{e:NavierBCsAlpha}
    v \cdot \mathbf{n} = 0 \text{ and }
    2D(v) \mathbf{n} \cdot \BoldTau
    + \al v \cdot \BoldTau = 0
    \text{ on } \Gamma,
\end{align}
where $\al$ is in $L^\iny(\Gamma)$, $\mathbf{n}$ and $\BoldTau$ are
unit normal and tangent vectors, respectively, to $\Gamma$, and $D(v)$
is the rate-of-strain tensor,
\begin{align*}
    D(v) = \frac{1}{2} \brac{\grad v + (\grad v)^T}.
\end{align*}
We follow the convention that $\mathbf{n}$ is an outward normal vector
and that the ordered pair $(\mathbf{n}, \BoldTau)$ gives the standard
orientation to $\R^2$. (We give an equivalent form of Navier boundary
conditions in \refC{BoundaryVorticity1}.)

J.L. Lions in \cite{JL1969} p. 87-98 and P.L. Lions in \cite{L1996} p.
129-131 consider the following boundary conditions, which we call
\textit{Lions} boundary conditions:
\[
    v \cdot \mathbf{n} = 0 \text{ and }
    \omega(v) = 0 \text{ on } \Gamma,
\]
where $\omega(v) =  \prt_1 v^2 - \prt_2 v^1$ is the vorticity of $v$.
Lions boundary conditions are the special case of Navier boundary
conditions in which $\al = 2 \kappa$, as we show in
\refC{LionsAsSpecialCase}.

J.L. Lions, in Theorem 6.10 p. 88 of \cite{JL1969}, proves existence
and uniqueness of a solution to the Navier-Stokes equations in the
special case of Lions boundary conditions, but includes the assumption
that the initial vorticity is bounded. With the same assumption of
bounded initial vorticity, the existence and uniqueness is established
in Theorem 4.1 of \cite{CMR} for Navier boundary conditions, under the
restriction that $\al$ is positive (and in $C^2(\Gamma)$). This is the
usual restriction, which is imposed to insure the conservation of
energy. Mathematically, negative values of $\al$ present no real
difficulty, so we do not make that restriction (until the last
section). The only clear gain from removing the restriction, however,
is that it allows us to view Lions boundary conditions as a special
case of Navier boundary conditions for more than just convex domains
(nonnegative curvature).

P.L. Lions establishes an energy inequality on p. 130 of \cite{L1996}
that can be used in place of the usual one for no-slip boundary
conditions. He argues that existence and uniqueness can then be
established---with no assumption on the initial vorticity---exactly as
was done for no-slip boundary conditions in the earlier sections of his
text. As we will show, P.L. Lions's energy inequality applies to Navier
boundary conditions in general, which gives us the same existence and
uniqueness theorem as for no-slip boundary conditions. (P.L. Lions's
comment on the regularity of $\pdx{u}{t}$ does not follow as in
\cite{L1996}, though, because (4.18) of \cite{L1996} is not valid for
general Navier boundary conditions.) Another method of proof is to
modify in a straightforward manner the classical proofs as they appear
in \cite{JL1969} and \cite{T2001}. In \refS{Existence} we state the
resulting existence and uniqueness theorem, but only prove the
corresponding energy bound, which we will need later. In
\refS{AdditionalRegularity}, we extend the existence, uniqueness,
regularity, and convergence results of \cite{CMR} and \cite{FLP} to
non-simply connected domains.

It is shown in \cite{FLP} that if the initial vorticity is in
$L^p(\Omega)$ for some $p > 2$, then after extracting a subsequence,
solutions to the Navier-Stokes equations with Navier boundary
conditions converge in $L^\iny([0, T]; L^2(\Omega))$ to a solution to
the Euler equations (with the usual boundary condition of tangential
velocity on the boundary) as $\nu \to 0$. This extends a result in
\cite{CMR} for initial vorticity in $L^\iny(\Omega)$, and because the
solution to the Euler equations is unique in this case, it follows that
the convergence is strong in $L^\iny([0, T]; L^2(\Omega))$---that is,
does not require the extraction of a subsequence.

The convergence in \cite{FLP} also generalizes the similar convergence
established for the special case of Lions boundary conditions on p. 131
of \cite{L1996} (though not including the case $p = 2$). The main
difficulty faced in making this generalization is establishing a bound
on the $L^p$-norms of the vorticity, a task that is much easier for
Lions boundary conditions (see p. 91-92 of \cite{JL1969} or p. 131 of
\cite{L1996}). In contrast, nearly all of \cite{CMR} and \cite{FLP},
including the structure of the existence proofs, is directed toward
establishing an analogous bound.

The methods of proof in \cite{CMR} and \cite{FLP} do not yield a bound
on the rate of convergence. With the assumptions in \cite{FLP}, such a
bound is probably not possible. We can, however, make an assumption
that is weaker than that of \cite{CMR} but stronger than that of
\cite{FLP} and achieve a bound on the rate of convergence.
Specifically, we assume, as in \cite{Y1995} and \cite{K2003}, that the
$L^p$-norms of the initial vorticity grow sufficiently slowly with $p$
(\refD{YudovichVorticity}) and establish the bound given in
\refT{VanishingViscosity}. To achieve this result, we also assume
additional regularity on $\al$ and $\Gamma$.

The bound on the convergence rate in $L^\iny([0, T]; L^2(\Omega))$ in
\refT{VanishingViscosity} is the same as that obtained for $\Omega =
\R^2$ in \cite{K2003}. In particular, it gives a bound on the rate of
convergence for initial vorticity in $L^\iny(\Omega)$ proportional to
\[
    (\nu t)^{\frac{1}{2}\exp\pr{-C\smallnorm{\omega^0}_{L^2 \cap L^\iny}t}},
\]
where $C$ is a constant depending on $\Omega$ and $\al$, and $\omega^0$
is the initial vorticity. This is essentially the same bound on the
convergence rate as that for $\Omega = \R^2$ appearing in \cite{C1996}.

Another interesting question is whether solutions to the
Navier-Stokes equations with Navier boundary conditions converge
to a solution to the Navier-Stokes equations with the usual
no-slip boundary conditions if we let the function $\al$ grow
large. We show in \refS{NoSlip} that such convergence does take
place for initial velocity in $H^3(\Omega)$ and $\Gamma$ in $C^3$
when we let $\al$ approach $+\iny$ uniformly on $\Gamma$. This
type of convergence is, in a sense, an inverse of the derivation
of the Navier boundary conditions from no-slip boundary conditions
for rough boundaries discussed in \cite{JM2001} and \cite{JM2003}.

We follow the convention that $C$ is always an unspecified constant
that may vary from expression to expression, even across an inequality
(but not across an equality). When we wish to emphasize that a constant
depends, at least in part, upon the parameters $x_1, \dots, x_n$, we
write $C(x_1, \dots, x_n)$. When we need to distinguish between
unspecified constants, we use $C$ and $C'$.

For vectors $u$ and $v$ in $\R^2$, we alternately write $\grad v u$ and
$u \cdot \grad v$, by both of which mean $u^i \prt_i v^j \mathbf{e}_j$,
where $\mathbf{e}_1, \mathbf{e}_2$ are basis vectors, and we define
$\grad u \cdot \grad v = u^{ij} v^{ij}$. Here, as everywhere in this
paper, we follow the common summation convention that repeated indices
are summed---whether or not one is a superscript and one a subscript.

If $X$ is a function space and $k$ a positive integer, we define
$(X)^k$ to be
\[
    \set{(f_1,\dots,f_k): f_1 \in X,\dots,f_k \in X}.
\]
For instance, $(H^1(\Omega))^2$ is the set of all vector fields, each
of whose components lies in $H^1(\Omega)$. To avoid excess notation,
however, we always suppress the superscript $k$ when it is clear from
the context whether we are dealing with scalar-, vector-, or
tensor-valued functions.

%
%
\section{Function Spaces}\label{S:FunctionSpaces}

\noindent Let
\begin{align}\label{e:ESpace}
    E(\Omega) = \set{v \in (L^2(\Omega))^2: \dv v \in L^2(\Omega)},
\end{align}
as in \cite{T2001}, with the inner product,
\[
    (u, v)_{E(\Omega)}
        = (u, v) + (\dv u, \dv v).
\]
We will use several times the following theorem, which is Theorem
1.2 p. 7 of \cite{T2001}.
\begin{lemma}\label{L:TraceThm}
    There exists a continuous linear operator
    $\gamma_{\mathbf{n}}$ mapping $E(\Omega)$ into $H^{-1/2}(\Gamma)$
    such that
    \begin{align*}
        \gamma_{\mathbf{n}}v =
            \text{ the restriction of } v \cdot \mathbf{n}
            \text{ to } \Gamma,
            \text{ for every } v \text{ in } (\Cal{D}(\ol{\Omega}))^2.
    \end{align*}

    Also, the following form of the divergence theorem is true for all
    vector fields $v$ in $E(\Omega)$ and scalar functions $h$ in $H^1(\Omega)$:
    \begin{align*}
        \int_\Omega v \cdot \grad h +
            \int_\Omega (\dv v)  h
                = \int_{\Gamma} \gamma_{\mathbf{n}} v
                    \cdot \gamma_0 h.
    \end{align*}
\end{lemma}

We always suppress the trace function $\gamma_0$ in our expressions,
and we write $v \cdot \mathbf{n}$ in place of $\gamma_{\mathbf{n}}v$.

Define the following function spaces as in \cite{CMR}:
\begin{align}
    \begin{split}\label{LionsNavierSpaces}
        H &= \set{v \in (L^2(\Omega))^2: \dv v = 0 \text{ in } \Omega
                \text{ and } v \cdot \mathbf{n} = 0
                \text{ on } \Gamma}, \\
        V &= \set{v \in (H^1(\Omega))^2: \dv v = 0 \text{ in } \Omega
                \text{ and } v \cdot \mathbf{n} = 0
                \text{ on } \Gamma}, \\
        \Cal{W} &= \set{v \in V \cap H^2(\Omega): v
                \text{ satisfies }
                \refE{NavierBCsAlpha}}.
    \end{split}
\end{align}
We give $\Cal{W}$ the $H^2$-norm, $H$ the $L^2$-inner product and norm,
which we symbolize by $(\cdot, \cdot)$ and
$\norm{\cdot}_{L^2(\Omega)}$, and $V$ the $H^1$-inner product,
\[
    (u, v)_V = \sum_i (\prt_i u, \prt_i v),
\]
and associated norm. This norm is equivalent to the $H^1$-norm, because
Poincar\'{e}'s inequality,
\begin{align}\label{e:Poincare}
    \norm{v}_{L^p(\Omega)}
        \le C(\Omega, p) \norm{\grad v}_{L^p(\Omega)}
\end{align}
for all $p$ in $[1, \iny]$, holds for all $v$ in $V$.

Ladyzhenskaya's inequality,
\begin{align}\label{e:Ladyzhenskaya}
    \norm{v}_{L^4(\Omega)}
        \le C(\Omega)
            \norm{v}_{L^2(\Omega)}^{1/2}
            \norm{\grad v}_{L^2(\Omega)}^{1/2}
\end{align}
also holds for all $v$ in $V$, though the constant in the inequality is
domain dependent, unlike the constant for the classical space $V$.

We will also frequently use the following inequality, which follows
from the standard trace theorem, Sobolev interpolation, and
Poincar\'{e}'s inequality:
\begin{align}\label{e:BoundaryNormBound}
    \norm{v}_{L^2(\Gamma)}
        \le C(\Omega) \norm{v}_{L^2(\Omega)}^{1/2}
            \norm{\grad v}_{L^2(\Omega)}^{1/2}
        \le C(\Omega) \norm{v}_V
\end{align}
for all $v$ in $V$.

%
%
\section{Hodge Decomposition of
$H$}\label{S:NonSimplyConnectedDomains}

\noindent Only simply connected domains are considered in
\cite{CMR} and \cite{FLP}. To handle non-simply connected domains
we will need a portion of the Hodge decomposition of
$L^2(\Omega)$. We briefly summarize the pertinent facts, drawing
mostly from Appendix I of \cite{T2001}.

Let $\Sigma_1, \dots, \Sigma_N$ be one-manifolds with boundary that
generate $H_1(\Omega, \Gamma; \R)$, the one-dimensional real homology
class of $\Omega$ relative to its boundary $\Gamma$.

We can decompose the space $H$ into two subspaces, $H = H_0 \oplus
H_c$, where
\begin{align*}
    H_0 &= \set{v \in H: \text{all internal fluxes are zero}}, \\
    H_c &= \set{v \in H: \omega(v) = 0}.
\end{align*}
An internal flux is a value of $\int_{\Sigma_i} v \cdot \mathbf{n}$.
Then $H_0 = H_c^\perp$ and there is an orthonormal basis ${\grad
q_1,\dots,\grad q_N}$ for $H_c \subseteq C^\iny(\ol{\Omega})$
consisting of the gradients of $N$ harmonic functions, $q_1,\dots,q_N$.
(Each $q_i$ is multi-valued in $\Omega$, but $\grad q_i$ is
single-valued.)

If $v$ is in $V$, then $v$ is also in $H$ so there exists a unique
$v_0$ in $H_0$ and $v_c$ in $H_c$ such that $v = v_0 + v_c$; also,
$(v_0, v_c) = 0$. But $v_c$ is in $C^\iny(\ol{\Omega})$ and so in $V$;
hence, $v_0$ also lies in $V$. This shows that $V = (V \cap H_0) \oplus
H_c$, though this is not an orthogonal decomposition of $V$.

The following is a result of Yudovich's:
\begin{lemma}\label{L:GradVsCurlV0}
    For any $p$ in $[2, \iny)$ and any $v$ in $V \cap H_0$,
    \begin{align*}
        \norm{\grad v}_{L^p(\Omega)}
            \le C(\Omega) p
                \norm{\omega(v)}_{L^p(\Omega)}.
    \end{align*}
\end{lemma}
\begin{proof}
    Let $v$ be in $V \cap H_0$. Since $v$ has no harmonic component, $v
    = \grad^\perp \psi = (-\prt_2 \psi, \prt_1 \psi)$ for some stream
    function $\psi$, which we can assume vanishes on $\Gamma$. Applying
    Corollary 1 of \cite{Y1962} with the operator $L = \Delta$ and $r
    = 0$ gives
    \begin{align*}
        \norm{\grad v}_{L^p(\Omega)}
            \le \norm{\psi}_{H^{2, p}(\Omega)}
            \le C(\Omega) p \norm{\Delta \psi}_{L^p(\Omega)}
            = C(\Omega) p \norm{\omega(v)}_{L^p(\Omega)}.
    \end{align*}
\end{proof}

For $\Omega$ simply connected, $H = H_0$, and \refL{GradVsCurlV0}
applies to all of $V$.
\begin{cor}\label{C:GradVsCurlV}
    For any $p$ in $[2, \iny)$ and any $v$ in $V$,
    \begin{align*}
        \norm{\grad v}_{L^p(\Omega)}
            \le C(\Omega) p
                \norm{\omega(v)}_{L^p(\Omega)}
                + C'(\Omega) \norm{v}_{L^2(\Omega)},
    \end{align*}
    the constants $C(\Omega)$ and $C'(\Omega)$ being independent of $p$.
\end{cor}
\begin{proof}
    Let $v$ be in $V$ with $v = v_0 + v_c$, where $v_0$ is
    in $V \cap H_0$ and $v_c$ is in $H_c$, and assume that $\grad v$ is in
    $L^p(\Omega)$.
    Let $v_c = \sum_{i=1}^N c_i \grad q_i$ and
    $r = \norm{v_c}_{L^2(\Omega)} = (\sum_i c_i^2)^{1/2}$. Then
    \begin{align*}
        \norm{\grad v_c}_{L^p(\Omega)}
           &= \sum_{i=1}^N \abs{c_i}
                    \norm{\grad \grad q_i}_{L^p(\Omega)}
            \le \sum_{i=1}^N r
                    \abs{\Omega}^{1/p}
                    \norm{\grad \grad q_i}_{L^\iny(\Omega)} \\
           &\le r \max\set{1, \abs{\Omega}^{1/2}} \sum_{i=1}^N
                    \norm{\grad \grad q_i}_{L^\iny(\Omega)}
            \le C \norm{v_c}_{L^2(\Omega)},
    \end{align*}
    where we used the smoothness of $\grad q_i$. But, $H_0 =
    H_c^\perp$, so $\norm{v}_{L^2(\Omega)} =
    \norm{v_0}_{L^2(\Omega)} + \norm{v_c}_{L^2(\Omega)}$ and
    thus $\norm{v_c}_{L^2(\Omega)} \le \norm{v}_{L^2(\Omega)}$.
    Therefore,
    \begin{align*}
        \norm{\grad v}_{L^p(\Omega)}
           &\le \norm{\grad v_0}_{L^p(\Omega)}
                + \norm{\grad v_c}_{L^p(\Omega)} \\
           &\le C(\Omega) p
                \norm{\omega(v)}_{L^p(\Omega)}
                + C'(\Omega) \norm{v}_{L^2(\Omega)}
    \end{align*}
    by virtue of \refL{GradVsCurlV0}.
\end{proof}

%
%
\section{Vorticity on the Boundary}\label{S:BoundaryVorticity}

\noindent If we parameterize each component of $\Gamma$ by arc length,
$s$, it follows that
\begin{align*}
    \pdx{\mathbf{n}}{\BoldTau}
        := \diff{\mathbf{n}}{s} = \kappa \BoldTau,
\end{align*}
where $\kappa$, the curvature of $\Gamma$, is continuous because
$\Gamma$ is $C^2$.

The second part of the following theorem is Lemma 2.1 of \cite{CMR},
and the first part is established similarly.

\begin{lemma}\label{L:BoundaryVorticity}
    If $v$ is in $(H^2(\Omega))^2$ with $v \cdot \mathbf{n} = 0$ on $\Gamma$,
    then
    \begin{align}\label{e:GradIdentity}
        \grad v \mathbf{n} &\cdot \BoldTau
            = \omega(v) + \grad v \BoldTau \cdot \mathbf{n}
            = \omega(v) - \kappa v \cdot \BoldTau,
    \end{align}
    and
    \begin{align}\label{e:StrainIdentity}
        D(v) \mathbf{n} \cdot \BoldTau
            = \frac{1}{2} \omega(v) - \kappa v \cdot \BoldTau.
    \end{align}
\end{lemma}

\begin{cor}\label{C:BoundaryVorticity1}
    A vector $v$ in $V \cap H^2(\Omega)$ satisfies Navier boundary
    conditions (that is, lies in $\Cal{W}$) if and only if
    \begin{align}\label{e:OmegaOnGamma}
        \omega(v) = (2 \kappa - \al) v \cdot \BoldTau
            \text{ and } v \cdot \mathbf{n} = 0
            \text{ on } \Gamma.
    \end{align}

    Also, for all $v$ in $\Cal{W}$ and $u$ in $V$,
    \begin{align}\label{e:GradBound}
        \grad v \mathbf{n} \cdot u
            = (\kappa - \al) v \cdot u
            \text{ on } \Gamma.
    \end{align}
\end{cor}
\begin{proof}
    Let $v$ be in $V \cap H^2(\Omega)$. Then from \refE{StrainIdentity},
    \begin{align}\label{e:IntBC}
        2D(v) \mathbf{n} \cdot \BoldTau
                + 2 \kappa(v \cdot \BoldTau)
            = \omega(v).
    \end{align}
    If $v$ satisfies Navier boundary conditions, then
    \refE{OmegaOnGamma} follows by subtracting
    $2D(v) \mathbf{n} \cdot \BoldTau
    + \al v \cdot \BoldTau = 0$ from \refE{IntBC}.
    Conversely, substituting the expression for $\omega(v)$ in
    \refE{OmegaOnGamma} into \refE{IntBC} gives $2D(v) \mathbf{n} \cdot \BoldTau
    + \al v \cdot \BoldTau = 0$.

    If $v$ is in $\Cal{W}$, then from \refE{GradIdentity},
    \begin{align*}
        \grad v \mathbf{n} \cdot \BoldTau
           &= \omega(v) - \kappa v \cdot \BoldTau
            = (2 \kappa - \al) v \cdot \BoldTau
                 - \kappa v \cdot \BoldTau
            = (\kappa - \al) v \cdot \BoldTau,
    \end{align*}
    and \refE{GradBound} follows from this, since $u$ is parallel to
    $\tau$ on $\Gamma$.
\end{proof}

\begin{cor}\label{C:LionsAsSpecialCase}
    For initial velocity in $H^2(\Omega)$,
    Lions boundary conditions are the special case of Navier
    boundary conditions where
    \[
        \al = 2 \kappa.
    \]
    That is, any solution of ($NS$) with Navier boundary conditions where $\al = 2
    \kappa$ is also a solution to ($NS$) with Lions boundary
    conditions.
\end{cor}

%
%
\section{Weak Formulation}\label{S:WeakFormulation}

\noindent For all $u$ in $\Cal{W}$ and $v$ in $V$,
\begin{align}\label{e:IntEq}
    \begin{split}
        \int_\Omega &\Delta u \cdot v
                    = \int_\Omega (\dv \grad u^i) v^i
        = \int_\Gamma (\grad u^i \cdot \mathbf{n}) v^i
                - \int_\Omega \grad u^i \cdot \grad v^i \\
        &= \int_\Gamma (\grad u \mathbf{n}) \cdot v
                - \int_\Omega \grad u \cdot \grad v
         = \int_\Gamma (\kappa - \al) u \cdot v
                - \int_\Omega \grad u \cdot \grad v,
    \end{split}
\end{align}
where we used \refE{GradBound} of \refC{BoundaryVorticity1}. This
motivates our formulation of a weak solution, in analogy with Problem
3.1 p. 190-191 of \cite{T2001}.

\begin{definition}\label{D:WeakSolution1}
    Given a viscosity $\nu > 0$ and initial velocity $u^0$ in $H$, $u$
    in $L^2([0, T]; V)$ is a weak solution to the Navier-Stokes equations
    (without forcing) if $u(0) = u^0$ and
    \begin{align*}
        \mathbf{(NS)}
        \qquad\diff{}{t} \int_\Omega u \cdot v
            + \int_\Omega (u \cdot \grad u) \cdot v
            + \nu \int_\Omega \grad u \cdot \grad v
            - \nu\int_\Gamma (\kappa - \al) u \cdot v
            = 0
    \end{align*}
    for all $v$ in $V$. (We make sense of
    the initial condition $u(0) = u^0$ as in \cite{T2001}.)
\end{definition}

Our formulation of a weak solution is equivalent to that in (2.11) and
(2.12) of \cite{CMR}. This follows from the identity,
\begin{align*}  
    2 \int_{\Omega} D(u) \cdot D(v)
        = \int_\Omega \grad u \cdot \grad v
            - \int_\Gamma \kappa u \cdot v,
\end{align*}
which holds for all $u$ and $v$ in $V$. This identity can be derived
from \refE{GradIdentity} and \refL{TraceThm}, and the density of
$H^2(\Omega) \cap V$ in $V$.

%
%
\section{Existence and Uniqueness}\label{S:Existence}

\noindent We prove only the energy bound of the following
existence and uniqueness theorem (see the comment in
\refS{Introduction}). We observe, however, that Ladyzhenskaya's
inequality, \refE{Ladyzhenskaya}, is required in the proof of
uniqueness.

\begin{theorem}\label{T:ExistenceAndUniqueness}
    Assume that $\Gamma$ is $C^2$ and $\al$ is in $L^\iny(\Gamma)$.
    Let $u^0$ be in $H$ and let $T > 0$.
    Then there exists a solution $u$ to ($NS$).
    Moreover, $u$ is in $L^2([0, T]; V) \cap C([0, T]; H)$,
    $u'$ is in $L^2([0, T]; V')$, and we have the energy inequality,
    \begin{align}\label{e:EnergyBound}
        \norm{u(t)}_{L^2(\Omega)}
            \le e^{C(\al) \nu t} \smallnorm{u^0}_{L^2(\Omega)},
    \end{align}
    where the constant $C(\al) = 0$ if $\al$ is nonnegative on
    $\Gamma$.
\end{theorem}
\begin{proof}
    We prove only \refE{EnergyBound}.
    We proceed with a Galerkin approximation as in the proof of Theorem
    3.1 on p. 192-193 of \cite{T2001}, but use the basis of
    \refC{ExistenceOfABasis}.
    Because this basis is also a basis for $H$, if we let $u^{0m}$ be
    the projection in $H$ of $u^0$ onto the span of the first $m$ basis vectors,
    then $u^{0m} \to u^0$ in $L^2(\Omega)$.
    Because the basis is in
    $H^2(\Omega)$, the approximate solution $u_m$ is in
    $C^1([0, T]; H^2(\Omega))$.

    \refD{WeakSolution1} leads to the following replacement for (3.27) p. 193 of
    \cite{T2001}:
    \begin{align*}
        (u_m'(t), u_m(t))
            + \nu \norm{\grad u_m(t)}_{L^2(\Omega)}^2
            = \nu\int_\Gamma (\kappa - \al) u_m \cdot u_m.
    \end{align*}
    Using \refE{GradBound} of \refC{BoundaryVorticity1} and Lemma 1.2 p. 176 of
    \cite{T2001}, we conclude that
    \begin{align}\label{e:PrePreEnergyBound}
        \frac{1}{2}\diff{}{t} \norm{u_m}_{L^2(\Omega)}^2
            + \nu \norm{\grad u_m}_{L^2(\Omega)}^2
            \le C \nu \norm{u_m}_{L^2(\Gamma)}^2,
    \end{align}
    where $C = \sup_\Gamma \abs{\kappa - \al}$.
    Except for the value of the constant, \refE{PrePreEnergyBound}
    is identical to the first inequality on p. 130 of \cite{L1996}, which
    is for the special case of Lions boundary conditions.

    Arguing exactly as in \cite{L1996}, it follows that
    \begin{align*}
        \diff{}{t} \norm{u_m}_{L^2(\Omega)}^2
            + \nu \norm{\grad u_m}_{L^2(\Omega)}^2
            \le C \nu \norm{u_m}_{L^2(\Omega)}^2.
    \end{align*}
    Integrating over time gives
    \begin{align}\label{e:PreEnergyBound}
        \begin{split}
            \norm{u_m(t)}_{L^2(\Omega)}^2
                &+ \nu \int_0^t \norm{\grad u_m(s)}_{L^2(\Omega)}^2 \, ds
                        \\
            &\le \smallnorm{u^{0m}}_{L^2(\Omega)}^2
                    +  C \nu \int_0^t \norm{u_m(s)}_{L^2(\Omega)}^2 \, ds.
        \end{split}
    \end{align}
    The energy bound,
    \begin{align}\label{e:EnergyBoundum}
        \norm{u_m(t)}_{L^2(\Omega)}^2
            \le e^{C \nu t} \smallnorm{u^{0m}}_{L^2(\Omega)}^2
            \le e^{C \nu t} \smallnorm{u^0}_{L^2(\Omega)}^2,
    \end{align}
    then follows from Gronwall's lemma,
    and shows
    that the right side of \refE{PreEnergyBound} is
    bounded uniformly in $[0, T]$. We conclude from
    \refE{PreEnergyBound} and \refE{EnergyBoundum} that
    \begin{align*}
        \set{u_m} \text{ is bounded in }
        L^2([0, T]; V) \cap L^\iny([0, T]; H),
    \end{align*}
    from which \refE{EnergyBound} will follow.
    (If $\al$ is nonnegative, then, in fact, energy is conserved---in the
    absence of forcing---so $C(\al) = 0$. This
    follows from the equation preceding (2.16) of \cite{CMR}.)
\end{proof}

%
%
\section{Additional Regularity}\label{S:AdditionalRegularity}

\noindent In this section we establish an existence theorem suited to
addressing the issue of convergence of a solution to ($NS$) to a
solution to the Euler equations, where we always impose stronger
regularity on the initial velocity.

If we assume extra regularity on the initial velocity, that regularity
will be maintained for all time. Our proof of this is an adaptation of
the proof of Theorem 3.5 p. 202-204 of \cite{T2001} to establish the
regularity of $u'$, combined with the second half of the proof of
Theorem 2.3 of \cite{CMR} to establish the regularity of $u$.

\begin{definition}\label{D:Compatible}
    A vector field $v$ in $\Cal{W}$ is called
    \textit{compatible} if $\omega(v)$ is in $L^\iny(\Omega)$.
\end{definition}
\refD{Compatible} is as in \cite{FLP}, except that we define the vector
field to be compatible instead of the vorticity.

\begin{theorem}\label{T:ExtraRegularity}
    Assume that $\Omega$ is a bounded domain with a
    $C^{2, 1/2 + \eps}$ boundary $\Gamma$ and that
    $\al$ is in $H^{1/2 + \eps}(\Gamma) + C^{1/2 + \eps}(\Gamma)$
    for some $\eps > 0$.
    Let $u^0$ be in $\Cal{W}$ with initial vorticity $\omega^0$, and let
    $u$ be the unique
    solution to ($NS$) given by \refT{ExistenceAndUniqueness} with
    corresponding vorticity $\omega$. Let $T > 0$. Then
    \[
        u' \in L^2([0, T]; V) \cap C([0, T]; H).
    \]
    If, in addition, $\omega^0$ is in $L^\iny(\Omega)$ (so $u^0$ is
    compatible), then
    \[
        u \in C([0, T]; H^2(\Omega)),\;
        \omega \in C([0, T]; H^1(\Omega)) \cap L^\iny([0, T] \times
        \ol{\Omega}).
    \]
\end{theorem}
\begin{proof}
    We prove the regularity of $u'$ in three steps as in the proof of
    Theorem 3.5 p. 202-204 of \cite{T2001}. The only change in step (i)
    is that we use the basis of \refC{ExistenceOfABasis}
    rather than the basis in \cite{T2001}.

    No change to step (ii) is required, because (3.88) of \cite{T2001}
    still holds.

    In step (iii), an additional term of
    \begin{align*}
        \nu\int_\Gamma (\kappa - \al) \smallabs{u'_m}^2
    \end{align*}
    appears on the right side of (3.94) of Temam's proof, which we
    bound by
    \begin{align*}
        C \nu \norm{u'_m}_{L^2(\Omega)}
              \norm{\grad u'_m}_{L^2(\Omega)}
           &\le \frac{\nu}{2} \norm{\grad u'_m}_{L^2(\Omega)}^2
              + C \nu \norm{u'_m}_{L^2(\Omega)}^2.
    \end{align*}
    Then (3.95) of Temam's proof becomes
    \begin{align*}
        \diff{}{t} \norm{u'_m(t)}_{L^2(\Omega)}^2
            \le \phi_m(t) \norm{u'_m(t)}_{L^2(\Omega)}^2,
    \end{align*}
    where
    \[
        \phi_m(t)
            = \pr{\frac{2}{\nu} + C \nu} \norm{u_m(t)}_{L^2(\Omega)}^2,
    \]
    and the proof of the regularity of $u'$ is completed as in
    \cite{T2001}, along with the observation in \cite{CMR} that
    $u'$ is then in $C([0, T];H)$.

    To prove the regularity of $u$ and $\omega$, we
    follow the argument in the second half of the proof of Theorem
    2.3 in \cite{CMR} (which does not rely on $\al$ being nonnegative).
    We must, however, impose
    additional regularity on $\Gamma$ and on $\al$
    over that assumed in \refT{ExistenceAndUniqueness}.
    This is to insure that $u$ lying in $C^{1/2}([0,
    T]; (H^1(\Omega))^2)$ implies that $(\kappa - \al/2) u \cdot \BoldTau$
    lies in $C^{1/2}([0, T]; H^1(\Omega))$. Our conditions on $\Gamma$
    and $\al$ are sufficient, though not necessary
    (see, for instance, Theorem 1.4.1.1 p. 21 and
    Theorem 1.4.4.2 p. 28 of \cite{G1985}).

    Then, after it is shown that $u$ is in $C([0, T]; (H^{2,
    q}(\Omega))^2)$, we know by Sobolev embedding that $u$ is in
    $C([0, T] \times \Omega)$. Thus,
    \[
        \norm{u \cdot \grad u (t)}_H
            \le \norm{u}_{L^\iny([0, T] \times \Omega)}
                \norm{u(t)}_V,
    \]
    and since we already have $u$ in $C([0, T]; V)$, it follows
    that $u \cdot \grad u$ and also $\Phi$ are in $C([0, T]; H)$. Then
    $\text{curl } \Phi$ is in $C([0, T]; H^{-1}(\Omega))$, and another
    pass through the argument in \cite{CMR}, this time with $q = 2$,
    gives $u$ in $C([0, T]; (H^2(\Omega))^2)$. Because the increase in
    regularity of the solution arises from the equation $-\grad \psi = w$ with the
    boundary condition $\psi = 0$, no regularity on $\Gamma$
    or on $\al$ beyond that we have assumed is required.

    (The argument in \cite{CMR} is for a simply connected domain.
    We can easily adapt it, though, by using the equivalent of
    Lemma 2.5 p. 26 of \cite{T2001}, which gives a stream
    function $\psi$ that is constant on each boundary component,
    which is good enough to apply Grisvard's result (Theorem 2.5.1.1 p.
    128 of \cite{G1985}) to conclude that $\psi$ is in $C([0, T]; H^{3,
    q}(\Omega))$.)
\end{proof}

With \refT{ExtraRegularity}, we have a replacement for Theorem 2.3 of
\cite{CMR} that applies regardless of the sign of $\al$. Since the
nonnegativity of $\al$ is used nowhere else in \cite{CMR} and
\cite{FLP}, all the results of both of those papers apply for simply
connected domains as well regardless of the sign of $\al$, but with the
extra regularity assumed on $\Gamma$ (and the lower regularity assumed
on $\al$).

To remove the restriction on the domain being simply connected, it
remains only to show that Lemmas 3.2 and 4.1 of \cite{FLP} remain valid
for non-simply connected domains. We show this for Lemma 3.2 of
\cite{FLP} in \refT{CompatibleSequence}. As for Lemma 4.1 of
\cite{FLP}, we need only use \refC{GradVsCurlV} to replace the term
$\norm{\omega(\cdot, t)}_{L^p(\Omega)}^{1 - \theta}$ with
$(\norm{\omega(\cdot, t)}_{L^p(\Omega)} + \norm{u(\cdot,
t)}_{L^2(\Omega)})^{1 - \theta}$ in the proof of Lemma 4.1 in
\cite{FLP}. Lemma 4.1 of \cite{FLP} then follows with no other changes
in the proof---only the value of the constant $C$ changes.

Let $u$ be the unique solution to ($NS$) given by Proposition 5.2 of
\cite{FLP}, and fix $q > 2$. By Lemma 4.1 of \cite{FLP} and
\refC{GradVsCurlV},
\begin{align}\label{e:LpBoundq}
    \begin{split}
        \norm{u}_{L^\iny([0, T]; V)}
            &= \norm{\grad u}_{L^\iny([0, T]; L^2(\Omega))}
            \le C\norm{\grad u}_{L^\iny([0, T]; L^q(\Omega))} \\
            &\le C(\norm{\omega}_{L^\iny([0, T]; L^q(\Omega))}
                + \norm{u}_{L^\iny([0, T]; L^2(\Omega))}) \\
            &\le C(T, \al, \kappa) e^{C(\al) \nu T}.
    \end{split}
\end{align}
Also, using Sobolev interpolation, \refE{Poincare}, and
\refC{GradVsCurlV},
\begin{align*}
    \begin{split}
        \norm{u(t)}_{C(\ol{\Omega})}
            &\le C \norm{u(t)}_{L^2(\Omega)}^\theta
                    \norm{u(t)}_{H^{1,q}(\Omega)}^{1 - \theta} \\
            &\le C\norm{u(t)}_{L^2(\Omega)}^\theta
                    (\norm{\omega(t)}_{L^q(\Omega)}
                    + \norm{u(t)}_{L^2(\Omega)})^{1 - \theta},
    \end{split}
\end{align*}
where $\theta = (q - 2)/(2q - 2)$. This norm is finite and bounded
over any finite range of viscosity by \refE{EnergyBound}. Using
Lemma 4.1 of \cite{FLP}, it follows that
\begin{align}\label{e:uInfBound}
    \norm{u}_{L^\iny([0, T] \times \Omega)} \le C
\end{align}
for all $\nu$ in $(0, 1]$, a bound we will use in
\refS{VanishingViscosity}.

%
%
\section{Vanishing Viscosity}\label{S:VanishingViscosity}

\noindent To describe Yudovich's conditions on the initial vorticity,
let $\phi: (1, \iny) \to [0, \iny)$ be any continuous function. We
define two functions, $\beta_{\eps, M, \phi}: [0, \iny) \to [0, \iny)$
and $\beta_{M, \phi}: [0, \iny) \to [0, \iny)$, parameterized by $\eps$
in $(0, 1)$, $M > 0$, and $\phi$:
\begin{align}\label{e:BetaExpression}
    \begin{split}
        &\beta_{\eps, M, \phi}(x)
            = M^\eps x^{1 - \eps} \phi(1/\eps), \\
        &\beta_{M, \phi}(x)
            = \inf \set{\beta_\eps(x): \eps \in (0, 1)}.
    \end{split}
\end{align}
For brevity, we write $\beta_\eps$ for $\beta_{\eps, M, \phi}$
and $\beta$ for $\beta_{M, \phi}$, with the choices of $M$ and $\phi$
being understood.

For all $\eps$ in $(0, 1)$, $\beta_\eps(x)$ is a monotonically
increasing function continuous in $x$ and in $\eps$, with $\lim_{x \to
0^+} \beta_\eps(x) = 0$. It follows that $\beta$ is a monotonically
increasing continuous function and that $\lim_{x \to 0^+} \beta(x) =
0$. Also, $\beta(x) \le \beta_\eps(x)$ for all $\eps$ in $(0, 1)$ and
$x \in [0, \iny)$.

\begin{definition}\label{D:Admissible}
    A continuous function $\theta:(1, \iny) \to [0, \iny)$ is called
    \textit{admissible} if
    \[
        \int_0^1 \frac{ds}{\beta_{M, \phi}(s)} = \iny,
    \]
    where $\phi(p) = p \theta(p)$. This condition is independent of the
    choice of $M$.
\end{definition}

Some examples of admissible functions are given in \cite{Y1995}.
Roughly speaking, a function is admissible if it does not grow much
faster than $\log p$.

\begin{definition}\label{D:YudovichVorticity}
    We say that a velocity vector $v$ has \textit{Yudovich vorticity} if $p
    \mapsto \norm{\omega(v)}_{L^p(\Omega)}$ is an admissible function.
\end{definition}

\begin{definition}\label{D:WeakSolutionEuler}
    Given an initial velocity $u^0$ in $V$, $u$
    in $L^2([0, T]; V)$ is a weak solution to the Euler equations
    if $u(0) = u^0$ and
    \begin{align*}
        \diff{}{t} \int_\Omega u \cdot v
            + \int_\Omega (u \cdot \grad u) \cdot v
            = 0
    \end{align*}
    for all $v$ in $V$.
\end{definition}

The existence of a weak solution to the Euler equations under the
assumption that the initial vorticity $\omega^0$ is in $L^p(\Omega)$
for some $p > 1$ (a weaker assumption than that of
\refD{WeakSolutionEuler} when $1 < p < 2$) was proved in \cite{Y1963}.
These solutions have the property that $\omega(u)$ is in
$L^\iny_{loc}(\R; L^p(\Omega))$. It is shown in \cite{Y1995} that
Yudovich initial vorticity is enough to insure uniqueness of solutions
for which $\omega(u)$ and $\prt_t u$ are in $L^\iny_{loc}(\R;
L^p(\Omega))$ for all $p$ in $[1, \iny)$. (Yudovich's uniqueness result
in \cite{Y1995} applies to a bounded domain in $\R^n$, although
existence is not known for $n > 2$. His approach works, with only very
minor changes, when applied to all of $\R^n$.)

In \cite{K2003}, it is shown that Yudovich initial vorticity is
sufficient to provide a bound on the rate of convergence in $L^\iny([0,
T]; L^2(\R^2))$ of solutions to the Navier-Stokes equations with
no-slip boundary conditions to the unique solution to the Euler
equations. In \refT{VanishingViscosity} we extend this result to
bounded domains when the Navier-Stokes equations have Navier boundary
conditions.

\begin{theorem}\label{T:VanishingViscosity}
    Assume that $\Omega$ and $\al$ are as in \refT{ExtraRegularity}.
    Fix $T > 0$ and let $u^0$ be in $V$
    and have Yudovich vorticity $\omega^0$.
    Let $\set{u_\nu}_{\nu > 0}$ be the solutions to ($NS$) given by
    5.2 of \cite{FLP} and $\ol{u}$ be the unique weak solution to the
    Euler equations for which $\omega(\ol{u})$ and $\prt_t \ol{u}$ are
    in $L^\iny_{loc}(\R;
    L^p(\Omega))$, $\ol{u}$ and each $u_\nu$ having initial velocity $u^0$.
    Then
    \[
        u_\nu(t) \to \ol{u}(t) \text{ in } L^\iny([0, T]; L^2(\Omega)
                \cap L^2(\Gamma))
           \text{ as } \nu \to 0.
    \]
    Also, there exists a constant $R = C(T, \al, \kappa)$, such that
    if we define the
    function $f:[0, \iny) \to [0, \iny)$ by
    \begin{align*}
        \int_{R \nu}^{f(\nu)} \frac{dr}{\beta(r)} = T,
    \end{align*}
    then
    \begin{align}\label{e:ConvergenceRate}
        \begin{split}
            \norm{u_\nu - \ol{u}}_{L^\iny([0, T]; L^2(\Omega))}
                    &\le f(\nu)
                \text{ and} \\
            \norm{u_\nu - \ol{u}}_{L^\iny([0, T]; L^2(\Gamma))}
                    &\le C'(T, \al, \kappa)\sqrt{f(\nu)}
        \end{split}
    \end{align}
    for all $\nu$ in $(0, 1]$.
\end{theorem}
\begin{proof}
    Let $w = u_\nu - \ol{u}$.
    It is possible to show that the integral identity in
    \refD{WeakSolution1} holds for any $v$ in $L^2([0, T]; V)$, as does
    the corresponding identity in \refD{WeakSolutionEuler}. Applying
    the resulting identities with $v = w$ and subtracting gives
    \begin{align}\label{e:OrigIntegral}
        \begin{split}
            \int_\Omega &w \cdot \prt_t w
                + \int_\Omega w \cdot(u_\nu \cdot \grad w)
                + \int_\Omega w \cdot (w \cdot \grad \ol{u}) \\
                &= \nu \int_\Gamma (\kappa - \al) u_\nu \cdot w
                - \nu \int_\Omega \grad u_\nu \cdot \grad w.
        \end{split}
    \end{align}

    Both $\prt_t u_\nu$ and $\prt_t \ol{u}$ are in $L^2([0, T]; V')$, so
    (see, for instance, Lemma 1.2 p. 176 of \cite{T2001}),
    \begin{align*}
        \int_\Omega &w \cdot \prt_t w
            = \frac{1}{2} \diff{}{t} \norm{w}_{L^2(\Omega)}^2.
    \end{align*}

    Applying \refL{TraceThm},
    \begin{align*}
        \int_\Omega &w \cdot(u_\nu \cdot \grad w) \\
           &= \int_\Omega w^i u_\nu^j \prt_j w^i
            = \frac{1}{2} \int_\Omega u_\nu^j
                \prt_j \sum_i (w^i)^2
            = \frac{1}{2} \int_\Omega u_\nu \cdot \grad \abs{w}^2 \\
           &= \frac{1}{2} \int_\Gamma (u_\nu \cdot \mathbf{n}) \abs{w}^2
                - \frac{1}{2} \int_\Omega (\dv u_\nu) \abs{w}^2
            = 0,
    \end{align*}
    since $u_\nu \cdot \mathbf{n} = 0$ on $\Gamma$
    and $\dv u_\nu = 0$ on $\Omega$. Thus,
    integrating \refE{OrigIntegral} over time,
    \begin{align}\label{e:IntegratedBound1}
        \norm{w(t)}_{L^2(\Omega)}^2
            \le A
                + 2 \int_0^t \int_\Omega \abs{w}^2 \abs{\grad
                    \ol{u}},
    \end{align}
    where
    \begin{align*}
        A
            = 2 \nu \int_0^t
                    \brac{
                            \int_\Gamma (\kappa - \al) u_\nu \cdot w
                          - \int_\Omega \grad u_\nu \cdot \grad w
                    }.
    \end{align*}

    Using \refE{BoundaryNormBound}, \refE{LpBoundq}, and the conservation
    of the $L^2$-norm of vorticity for the Euler equation,
    we have
    \begin{align}\label{e:kappaalphawuBound}
        \begin{split}
           &\abs{\int_\Gamma (\kappa - \al) u_\nu \cdot w}
                \le \norm{\kappa - \al}_{L^\iny(\Gamma)}
                    \norm{u_\nu \cdot w}_{L^1(\Gamma)} \\
               &\qquad\le \norm{\kappa - \al}_{L^\iny(\Gamma)}
                    \norm{\grad u_\nu}_{L^2(\Omega)}
                    \norm{\grad w}_{L^2(\Omega)}
                \le C(T, \al, \kappa)e^{C(\al)\nu T}.
        \end{split}
    \end{align}
    By \refE{LpBoundq} we also have
    \begin{align}\label{e:wuBound}
        \abs{\int_\Omega \grad u_\nu \cdot \grad w}
            \le \norm{\grad u_\nu}_{L^2(\Omega)}
                  \norm{\grad w}_{L^2(\Omega)}
            \le C(T, \al, \kappa)e^{C(\al)\nu T},
    \end{align}
    so $A \le C(T, \al, \kappa) e^{C(\al)\nu T} \nu$.

    By \refE{uInfBound},
    $\norm{u_\nu}_{L^\iny([0, T] \times
    \Omega)} \le C$ for all $\nu$ in $(0, 1]$.
    It is also true that $\ol{u}$ is in $L^\iny([0, T] \times
    \Omega)$ (arguing, for instance,
    exactly as in the derivation of \refE{uInfBound}). Thus,
    \[
        M = \sup_{\nu \in (0, 1]}
                \smallnorm{\abs{w}^2}_{L^\iny([0, T] \times \Omega)}
    \]
    is finite.

    Also, because
    vorticity is conserved for $\ol{u}$, we have, by \refC{GradVsCurlV},
    \begin{align}\label{e:phi}
        2 \norm{\grad \ol{u}(t)}_{L^p(\Omega)}
            \le Cp \smallnorm{\omega^0}_{L^p(\Omega)}
                + C\norm{\ol{u}}_{L^2(\Omega)}
            =: \phi(p)
    \end{align}
    for all $p \ge 2$.
    Then, as in \cite{K2003},
    \[
        2\int_\Omega \abs{w}^2 \abs{\grad \ol{u}}
            \le \beta(\LTwoNorm{w}^2),
    \]
    where $\beta = \beta_{M, \phi}$ is the function in
    \refE{BetaExpression}. (The additive constant
    $C\norm{\ol{u}}_{L^2(\Omega)}$ in \refE{phi}
    does not affect the integral condition in \refD{Admissible}.)

    Letting $L(t) = \norm{w(t)}_{L^2(\Omega)}^2$, we have
    \begin{align}\label{e:DrivingInequality}
        \begin{split}
            L(t)
                \le A
                + \int_0^t \beta(L(r)) \, dr.
        \end{split}
    \end{align}

    Using Osgood's lemma as in \cite{K2003}, we conclude that
    \begin{align}\label{e:IntL}
        \int_A^{L(t)} \frac{dr}{\beta(r)} \le t,
    \end{align}
    and that as $\nu \to 0$, $A \to 0$, and $L(t) \to 0$ uniformly
    over any finite time interval. The rate of convergence given in
    $L^\iny([0, T]; L^2(\Omega))$  in
    \refE{ConvergenceRate} can be
    derived from \refE{IntL} precisely as in \cite{K2003}.

    By \refE{BoundaryNormBound},
    \begin{align*}
        \norm{u_\nu - \ol{u}}_{L^2(\Gamma)}
           &= \norm{w}_{L^2(\Gamma)}
            \le C\norm{\grad w}_{L^2(\Omega)}^{1/2}
                 \norm{w}_{L^2(\Omega)}^{1/2} \\
           &\le C(T, \al, \kappa)e^{C(\al)\nu T} L(t)^{1/2},
    \end{align*}
    from which the convergence rate for $L^\iny([0, T]; L^2(\Gamma))$ in
    \refE{ConvergenceRate} follows.
\end{proof}

The convergence rate in $L^\iny([0, T]; L^2(\Omega))$ established in
\refT{VanishingViscosity} is the same as that established for the
entire plane in \cite{K2003}, except for the values of the constants.

%
%
\section{No-slip Boundary Conditions}\label{S:NoSlip}

\noindent As long as $\al$ is non-vanishing, we can reexpress the
Navier boundary conditions in \refE{NavierBCsAlpha} as
\begin{align}\label{e:NavierBCsGamma}
    v \cdot \mathbf{n} = 0 \text{ and }
    2\gamma D(v) \mathbf{n} \cdot \BoldTau
    + v \cdot \BoldTau = 0
    \text{ on } \Gamma,
\end{align}
where $\gamma = 1/\al$. When $\gamma$ is identically zero, we have the
usual no-slip boundary conditions. An obvious question to ask is
whether it is possible to arrange for $\gamma$ to approach zero in such
a manner that the corresponding solutions to the Navier-Stokes
equations with Navier boundary conditions approach the solution to the
Navier-Stokes equations with the usual no-slip boundary conditions in
$L^\iny([0, T]; L^2(\Omega))$.

Let $u^0$ be an initial velocity in $V$, and assume that $\gamma
> 0$ lies in $L^\iny(\Gamma)$. Fix a $\nu > 0$ and let
\begin{align*}
    \begin{array}{llll}
        u_{\nu,\gamma} &=&\text{the unique solution to the Navier-Stokes equations} \\
            &&\text{with Navier boundary conditions for } \al =
                    1/\gamma
              \text{ and}\\
        \widetilde{u}_\nu &=&\text{the unique solution to the Navier-Stokes equations} \\
            &&\text{with no-slip boundary conditions,} \\
    \end{array}
\end{align*}
in each case with the same initial velocity $u^0$. (In
\refT{VanishingViscosity} we wrote $u_{\nu,\gamma}$ as
$u_\nu$.)

If we let $\gamma$ approach 0 uniformly on the boundary, we
automatically have some control over $u_{\nu, \gamma}$ on the boundary.
\begin{lemma}\label{L:BoundaryBound}
    For sufficiently small $\norm{\gamma}_{L^\iny(\Gamma)}$,
    \begin{align}\label{e:AutomaticBoundaryVelocityBound}
        \norm{u_{\nu, \gamma}}_{L^2([0, T];
                L^2(\Gamma))}
            \le \frac{\smallnorm{u^0}_{L^2(\Omega)}}{\sqrt{\nu}}
                 \norm{\gamma}_{L^\iny(\Gamma)}^{1/2}.
    \end{align}
\end{lemma}
\begin{proof}
    Assume that $\norm{\gamma}_{L^\iny(\Gamma)}$ is sufficiently small
    that $\al > \kappa$ on
    $\Gamma$. Then, as in the proof of \refT{ExistenceAndUniqueness}, we
    have
    \begin{align*}
        \frac{1}{2}\diff{}{t} &\norm{u_{\nu, \gamma}(t)}_{L^2(\Omega)}^2
            + \nu \norm{\grad u_{\nu, \gamma}(t)}_{L^2(\Omega)}^2
            = \nu\int_\Gamma (\kappa - \al) u_{\nu, \gamma} \cdot
            u_{\nu, \gamma},
    \end{align*}
    so,
    \begin{align*}
        \norm{u_{\nu, \gamma}(t)}_{L^2(\Omega)}^2
           &\le \smallnorm{u^0}_{L^2(\Omega)}^2
           + 2 \nu \int_0^t \int_\Gamma (\kappa - \al) u_{\nu, \gamma}
                    \cdot u_{\nu, \gamma}.
    \end{align*}
    But,
    \begin{align*}
        \int_\Gamma (\kappa - \al) u_{\nu, \gamma}
                    \cdot u_{\nu, \gamma}
            \le
            - \inf_\Gamma \set{\al - \kappa}
                \norm{u_{\nu, \gamma}(t)}_{L^2(\Gamma)}^2,
    \end{align*}
    so
    \begin{align*}
        \norm{u_{\nu, \gamma}(t)}_{L^2(\Omega)}^2
           &\le \smallnorm{u^0}_{L^2(\Omega)}^2
              - 2\nu \inf_\Gamma \set{\al - \kappa}
                 \norm{u_{\nu, \gamma}}_{L^2([0, t]; L^2(\Gamma))}^2
    \end{align*}
    and
    \begin{align*}
        \norm{u_{\nu, \gamma}}_{L^2([0, t]; L^2(\Gamma))}^2
            \le \smallnorm{u^0}_{L^2(\Omega)}^2
                /(2\nu \inf_\Gamma \set{\al - \kappa}).
    \end{align*}
    Then \refE{AutomaticBoundaryVelocityBound} follows because
    $\norm{\gamma}_{L^\iny(\Gamma)}
    \inf_\Gamma \set{\al - \kappa} \to 1$ as
    $\norm{\gamma}_{L^\iny(\Gamma)} \to 0$.
\end{proof}

If we assume enough smoothness of the initial data and of $\Gamma$, we
can use \refE{AutomaticBoundaryVelocityBound} to establish convergence
of $u_{\nu, \gamma}$ to $\widetilde{u}_\nu$ as
$\norm{\gamma}_{L^\iny(\Gamma)} \to 0$.

\begin{theorem}\label{T:NavierToNoSlipSmooth}
    Fix $T > 0$, assume that $u^0$ is in $V \cap
    H^3(\Omega)$ with $u^0 = 0$ on $\Gamma$,
    and assume that $\Gamma$ is $C^3$.
    Then for any fixed $\nu > 0$,
    \begin{align}\label{e:NavierToNoSlipSmooth}
        u_{\nu, \gamma} \to \widetilde{u}_\nu \text{ in }
                L^\iny([0, T]; L^2(\Omega)) \cap L^2([0, T]; L^2(\Gamma))
    \end{align}
    as $\gamma \to 0$ in $L^\iny(\Gamma)$.
\end{theorem}
\begin{proof}
    First, $u_{\nu, \gamma}$ exists and is unique by
    \refT{ExistenceAndUniqueness};
    the existence and uniqueness of $\widetilde{u}_\nu$ is a
    classical result. Because $u^0$ is in $H^3(\Omega)$ and $\Gamma$ is $C^3$,
    $\widetilde{u}_\nu$ is in $L^\iny([0, T];
    H^3(\Omega))$ by the argument on p. 205 of \cite{T2001} following
    the proof of Theorem 3.6 of \cite{T2001}. Hence, $\grad \widetilde{u}_\nu$ is
    in $L^\iny([0, T]; H^2(\Omega))$ and so in $L^\iny([0, T]; C(\Omega))$.

    Arguing as in the proof of \refT{VanishingViscosity} with
    $w = u_{\nu, \gamma} - \widetilde{u}_\nu$, we have
    \begin{align*}
        \int_\Omega &\prt_t w \cdot w
            + \int_\Omega w \cdot (u_{\nu, \gamma} \cdot \grad w)
            + \int_\Omega w \cdot (w \cdot \grad \widetilde{u}_\nu)
            + \int_\Omega \grad w \cdot \grad w \\
           &\qquad- \nu \int_\Gamma (\kappa - \al) u_{\nu, \gamma} \cdot w
            + \nu \int_\Gamma (\grad \widetilde{u}_\nu \mathbf{n})
                \cdot w = 0.
    \end{align*}
    But $\widetilde{u}_\nu = 0$ on $\Gamma$ so $w = u_{\nu, \gamma}$ on
    $\Gamma$, and
    \begin{align*}
        \int_\Omega &\prt_t w \cdot w
            + \int_\Omega w \cdot (w \cdot \grad \widetilde{u}_\nu)
            + \int_\Omega \abs{\grad w}^2
            + \nu \int_\Gamma (\al - \kappa) \abs{u_{\nu, \gamma}}^2 \\
           &\qquad+ \nu \int_\Gamma (\grad \widetilde{u}_\nu \mathbf{n})
                \cdot u_{\nu, \gamma}
            = 0. \\
    \end{align*}

    Then, for $\norm{\gamma}_{L^\iny(\Gamma)}$ sufficiently small that
    $\al = 1/\gamma > \kappa$ on $\Gamma$,
    \begin{align}\label{e:IntegratedBound}
        \norm{w(t)}_{L^2(\Omega)}^2
            \le A
                + 2 \int_0^t \int_\Omega \abs{w}^2 \abs{\grad
                    \widetilde{u}_\nu},
    \end{align}
    where
    \begin{align*}
        A
            = -2 \nu \int_0^t \int_\Gamma (\grad \widetilde{u}_\nu \mathbf{n})
                \cdot u_{\nu, \gamma}.
    \end{align*}

    By \refE{GradIdentity}, $(\grad \widetilde{u}_\nu \mathbf{n}) \cdot
    \BoldTau = \omega(\widetilde{u}_\nu) - \kappa \widetilde{u}_\nu \cdot \BoldTau
    = \omega(\widetilde{u}_\nu)$ on
    $\Gamma$. But $u_{\nu, \gamma}$ is parallel to $\BoldTau$ on $\Gamma$, so
    $(\grad \widetilde{u}_\nu \mathbf{n}) \cdot u_{\nu, \gamma} =
    \omega(\widetilde{u}_\nu) u_{\nu, \gamma} \cdot \BoldTau$.
    Thus,
    \begin{align*}
        -\int_\Gamma (\grad \widetilde{u}_\nu \mathbf{n})
                \cdot u_{\nu, \gamma}
           &= -\int_\Gamma \omega(\widetilde{u}_\nu) u_{\nu, \gamma} \cdot \BoldTau
            \le \norm{\omega(\widetilde{u}_\nu)}_{L^2(\Gamma)}
                \norm{u_{\nu, \gamma} \cdot \BoldTau}_{L^2(\Gamma)} \\
           &\le C\norm{\widetilde{u}_\nu}_{H^2(\Omega)}
                \norm{u_{\nu, \gamma} \cdot \BoldTau}_{L^2(\Gamma)},
    \end{align*}
    so
    \begin{align*}
        A
           &\le C\nu \norm{\widetilde{u}_\nu}_{L^2([0, T]; H^2(\Omega))}
                  \norm{u_{\nu, \gamma}}_{L^2([0, T]; L^2(\Gamma))}.
    \end{align*}

    By Theorem 3.10 p. 213 of \cite{T2001},
    $\norm{\widetilde{u}_\nu}_{L^2([0, T]; H^2(\Omega))}$
    is finite (though the bound on it in \cite{T2001}
    increases to infinity as $\nu$ goes to 0), so by
    \refL{BoundaryBound},
    \begin{align}\label{e:ABound}
        A
            \le C_1(\nu) \norm{\gamma}_{L^\iny(\Gamma)}^{1/2}.
    \end{align}

    Because $\grad \widetilde{u}_\nu$ is in
    $L^\iny([0, T]; C(\ol{\Omega}))$,
    \begin{align*}
        \int_0^t \int_\Omega \abs{w}^2 \abs{\grad \widetilde{u}_\nu}
            \le C_2(\nu) \int_0^t \norm{w(s)}_{L^2(\Omega)}^2 \, ds,
    \end{align*}
    where $C_2(\nu) = \norm{\grad \widetilde{u}_\nu}_{L^\iny([0, T] \times \Omega)}$,
    and \refE{IntegratedBound} becomes
    \begin{align*}
        \norm{w(t)}_{L^2(\Omega)}^2
            \le
                C_1(\nu) \norm{\gamma}_{L^\iny(\Gamma)}^{1/2} +
                C_2(\nu) \int_0^t \norm{w(s)}_{L^2(\Omega)}^2 \, ds.
    \end{align*}
    By Gronwall's Lemma,
    \begin{align*}
        \norm{w(t)}_{L^2(\Omega)}^2
            \le C_1(\nu) \norm{\gamma}_{L^\iny(\Gamma)}^{1/2}
                    e^{C_2(\nu)t},
    \end{align*}
    and the convergence in $L^\iny([0, T]; L^2(\Omega))$ follows immediately.
    Convergence in $L^2([0, T]; L^2(\Gamma))$ follows directly from
    \refL{BoundaryBound}, since $\widetilde{u}_\nu = 0$ on $\Gamma$.
\end{proof}

We cannot prove convergence in $L^\iny([0, T]; L^2(\Gamma))$ as we did
in \refT{VanishingViscosity}, because we do not have a bound on the
vorticity of $u_{\nu, \gamma}$ that is uniform over sufficiently small
values of $\norm{\gamma}_{L^\iny(\Gamma)}$. But if we did have such a
bound, we could also establish convergence in $L^\iny([0, T];
L^2(\Omega) \cap L^2(\Gamma))$ when $u^0$ in $V \cap H^2(\Omega)$ has
Yudovich initial vorticity by combining the approaches in the proofs of
\refT{VanishingViscosity} and \refT{NavierToNoSlipSmooth}.

\appendix
%
%
\section{Compatible Sequences}\label{S:VariousLemmas}

\noindent For $p$ in $(1, \iny)$, define the spaces
\begin{align}\label{e:VpSpaces}
    X_0^p = H_0\cap H^{1, p}(\Omega)
        \text{ and }
    X^p = H\cap H^{1, p}(\Omega) = X_0^p \oplus H_c,
\end{align}
each with the $H^{1,p}(\Omega)$-norm.

\begin{lemma}\label{L:FLPInequality}
    Let $p$ be in $(1, \iny]$. For $p < 2$ let $\widehat{p} = p/(2 - p)$,
    for $p > 2$ let $\widehat{p} = \iny$, and for $p = 2$ let $\widehat{p}$ be
    any value in $[2, \iny]$. Then for any $v$ in $X_0^p$,
    \begin{align*}
        \norm{v}_{L^{\widehat{p}}(\Gamma)}
            \le C(p) \norm{\omega(v)}_{L^p(\Omega)}.
    \end{align*}
\end{lemma}
\begin{proof}
    For $p < 2$ and any $v$ in $X_0^p$, we have
    \begin{align*}
        \norm{v}_{L^{\widehat{p}}(\Gamma)}
           &\le C(p) \norm{v}_{L^p(\Omega)}^{1 - \lambda}
                     \norm{\grad v}_{L^p(\Omega)}^\lambda
            \le C(p) \norm{\grad v}_{L^p(\Omega)} \\
           &\le C(p) \norm{\omega(v)}_{L^p(\Omega)},
    \end{align*}
    where $\la = 2(\widehat{p} - p)/(p(\widehat{p} - 1)) = 1$ if $p < 2$ and $\la = 2/p$ if
    $p \ge 2$. The first inequality follows from Theorem 3.1 p. 42 of
    \cite{G1994}, the second follows from \refE{Poincare}, and the
    third from \refL{GradVsCurlV0}.
\end{proof}

Given a vorticity $\omega$ in $L^p(\Omega)$ with $p$ in $(1, \iny)$,
the Biot-Savart law gives a vector field $v$ in $H$ whose vorticity is
$\omega$. (That $v$ is in $L^2(\Omega)$ follows as in the proof of
\refL{FLPInequality}, $\Omega$ being bounded.) Let $v = v_0 + v_c$,
where $v_0$ is in $H_0$ and $v_c$ is in $H_c$. Then $\omega(v_0) =
\omega$ as well, so we can define a function $K_\Omega$: $L^p(\Omega)
\to H_0$ by $\omega \mapsto v_0$ having the property that
$\omega(K_\Omega(\omega)) = \omega$. By \refE{Poincare} and
\refL{GradVsCurlV0}, $v_0$ is also in $H^{1,p}(\Omega)$, so in fact,
$K_\Omega$: $L^p(\Omega) \to X_0^p$ and is the inverse of the function
$\omega$. It is continuous by the same two lemmas.

\begin{theorem}\label{T:CompatibleSequence}
    Assume that $\Gamma$ is $C^2$ and $\al$ is in $L^\iny(\Gamma)$.
    Let $\ol{v}$ be in $X^p$ for some
    $p$ in $(1, \iny)$ and have vorticity $\ol{\omega}$. Then there
    exists a sequence $\set{v_i}$ of compatible
    vector fields (\refD{Compatible}) whose
    vorticities converge strongly to $\ol{\omega}$ in $L^p(\Omega)$.
    The vector fields $\set{v_i}$ converge strongly to $\ol{v}$ in $X^p$
    and, if $p \ge 2$, also in $V$.
\end{theorem}
\begin{proof}
    We adapt the proof of Lemma 3.2 of \cite{FLP}.
    Suppose that $\ol{v} = \ol{v}_0 + \ol{v}_c$
    with $\ol{v}_0 \in X_0^p$
    and $\ol{v}_c$ in $H_c$. Define $\beta$ as in Equation (3.1) of
    \cite{FLP}, but let $v = K_\Omega[\beta] + \ol{v}_c$ and start the
    iteration with $\omega_1 = \ol{\omega}$.
    Then the fixed point argument goes through unchanged because $v_1 -
    v_2$ is in $X_0^p$ and we can apply \refL{FLPInequality}.
    The only further change is the estimate on
    $\norm{G^n}_{L^{\widehat{p}}(\Gamma)}$, which becomes
    \begin{align*}
        \norm{G^n}_{L^{\widehat{p}}(\Gamma)}
           &\le \norm{2 \kappa - \al}_{L^\iny}
                \norm{K_\Omega[\omega_n] + \ol{v}_c}_{L^{\widehat{p}}(\Gamma)}
                    \\
           &\le C_p (\norm{\omega}_{L^p(\Omega)} +
                     \norm{\ol{v}_c}_{L^{\widehat{p}}(\Gamma)})
                    + \frac{1}{2} \norm{G^n}_{L^{\widehat{p}}(\Gamma)},
    \end{align*}
    for $n$ sufficiently large, which is still sufficient to imply the
    required bound that insures convergence of $\omega_n$ to $\ol{\omega}$
    in $L^p(\Omega)$.

    Letting $v_n = K_\Omega[\omega_n] + \ol{v}_c$, we have
    \begin{align*}
        \norm{\grad \ol{v} - \grad v_n}_{L^p(\Omega)}
           &= \norm{\grad \ol{v}_0 + \grad \ol{v}_c
                    - (\grad K_\Omega[\omega_n] + \grad
                    \ol{v}_c)}_{L^p(\Omega)} \\
           &= \norm{\grad(\ol{v}_0 - K_\Omega[\omega_n])}_{L^p(\Omega)}
                   \\
           &\le Cp\norm{\omega(\ol{v}_0 - K_\Omega[\omega_n])}_{L^p(\Omega)}
            = Cp\norm{\ol{\omega} - \omega_n}_{L^p(\Omega)},
    \end{align*}
    where we used \refL{GradVsCurlV0}. Then by \refE{Poincare},
    $v_n$ converges strongly to $\ol{v}$ in $X^p$ as well.
    Convergence in $V$ for $p \ge 2$ follows since $\Omega$ is bounded.
\end{proof}

We only require \refT{CompatibleSequence} for $p \ge 2$. We
include all the cases, however, for the same reason as in
\cite{FLP}: in the hope that if the vorticity bound in Lemma 4.1
of \cite{FLP} can be extended to $p$ in $(1, 2)$, then the
convergence in Proposition 5.2 of \cite{FLP} can also be extended
(for non-simply connected $\Omega$).

\begin{cor}\label{C:ExistenceOfABasis}
    Assume that $\Gamma$ is $C^2$, and $\al$ is in $L^\iny(\Gamma)$.
    Then there exists a basis for $V$ lying in $\Cal{W}$  that is also
    a basis for $H$.
\end{cor}
\begin{proof}
    The space $V = (V \cap H_0) \oplus H_c$ is separable because
    $V \cap H_0$ is the image under the
    continuous function $K_\Omega$ of the separable space $L^2(\Omega)$ and
    $H_c$ is finite-dimensional. Let $\set{v_i}_{i=1}^\iny$ be a dense
    subset of $V$. Applying \refT{CompatibleSequence} to each $v_i$ and
    unioning all the sequences, we obtain a countable subset
    $\set{u_i}_{i=1}^\iny$ of $\Cal{W}$ that is dense in $V$. Selecting a maximal
    independent set gives us a basis for $V$ and for $H$ as well,
    since $V$ is dense in $H$.
\end{proof}

%
%
\section*{Acknowledgements}\label{S:Acknowledgements}

I wish to thank Josef M\'{a}lek for suggesting that I look at Navier
boundary conditions, and Misha Vishik for many useful discussions.

%
%

\end{document}